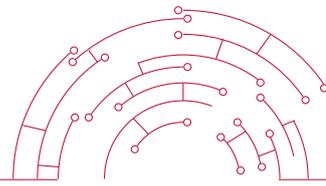

# Redefining Astronomy Summer Camps in the Age of the Pandemic: a Break from the IAYC's 50-Year History

Eva-Maria AHRER[*1,2], Melanie ARCHIPLEY[*1,3], Hannah S. DALGLEISH[*1,4] and Daniel J. MORTIMER[*1,5]

**Abstract.** The International Astronomical Youth Camp (IAYC) is a 50-year old summer camp, where participants work independently on astronomy projects. Due to the ongoing COVID-19 pandemic, the 2020 and 2021 instalments of the IAYC were cancelled, a first in the camp's history. An online format was established dubbed the eIAYC, consisting of three types of activities: (1) an engagement series with astronomical talks and workshops; (2) small independent research projects; and (3) a non-astronomical program involving a range of social activities. Here we present the experience of adapting an in-person camp into an online alternative in order to further the IAYC's mission. We discuss organisational challenges, experiences with online engagement, and how the 2020 eIAYC informed plans for this year's eIAYC.

## 1. Introduction

The COVID-19 pandemic has presented unprecedented challenges for all outreach organisations. The International Workshop for Astronomy e.V. (IWA) has been organising the three-week International Astronomical Youth Camp (IAYC) for more than 50 years, however in 2020, the traditional IAYC had to be cancelled. IWA thus found a virtual alternative to the in-person summer camp, overhauling the camp's structure to offer activities to engage with the IAYC community, both academically and culturally. With this in mind and constrained by limited resources, the eIAYC 2020 activities were restricted to former participants of the IAYC, as the foremost goal was to keep the community engaged with the organisation.

## 2. The eIAYC 2020

The online format of the IAYC, dubbed eIAYC, ran from 16 June until 30 August 2020 and had 63 participants overall. It consisted of three types of activities: an engagement series, project groups and a non-astronomical program.

*1 International Workshop for Astronomy e.V. (IWA)
*2 University of Warwick
   eva-maria@iayc.org
*3 University of Illinois Urbana-Champaign
   melanie@iayc.org
*4 University of Namibia
   hannah@iayc.org
*5 University of Cambridge
   daniel@iayc.org

### 2.1 Engagement Series

The engagement series consisted of astronomy-related talks and workshops every two to three weeks. Examples include a python programming and data visualization tutorial and a "How to build a telescope simulator" workshop. These events were usually presented via Google Meet and had around ten attendees per session.

### 2.2 Project Groups

This part of the eIAYC was constructed as an equivalent to the projects that participants would normally do during the three-week in-person instalments. Four members of IWA provided projects for the participants to work on over a few weeks during the summer months. Topics were astronomy-related and based on the respective research of the group leaders, including instrumentation and machine learning, astronomy for development, exoplanets and eclipsing binaries. The participants worked in groups of two to four, some of them alone, and up to eight per leader. Afterwards, the participants had the opportunity to present their work in the form of an online poster to other participants. An example of work completed during a project is shown in Fig. 1.

### 2.3 Non-Astronomical Program (NAP)

While putting on astronomy outreach events and offering research experience are key priorities for the IAYC, international and cultural exchange is just as integral to the mission. For this purpose, every participant was put into one of three different groups and challenges were issued, both as team and individual activities. The challenges had to be possible from home e.g. "What can you build from a single A4 sheet of paper?". Other examples for NAP included a pub quiz, a movie night and a performance evening.

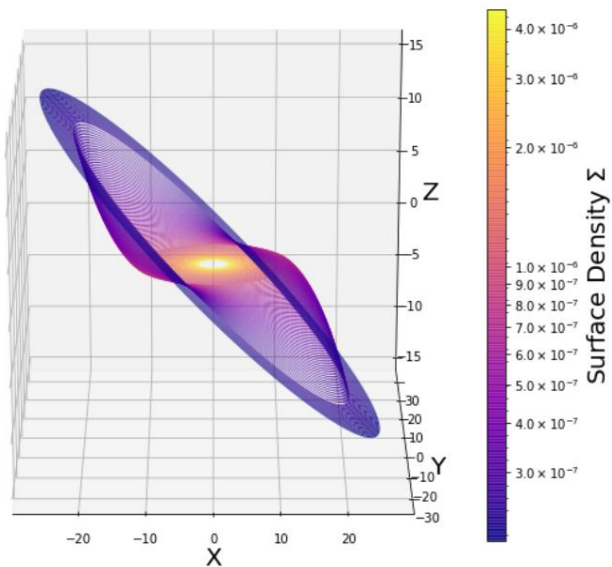

Fig. 1 A simulation of a warped protoplanetary disk. This was part of a project conducted within the eIAYC 2020 by Carolin Kimmig.

### 3. Organisational Challenges

As the IAYC community is spread across the globe [1], the eIAYC organisers had to determine the best times to offer activities. In the end, almost all participants and organisers were based in Europe and North America, so European evening times were chosen for all events.

To communicate efficiently and cost-effectively with the participants, the eIAYC used a Discord server as the main online platform. This allowed leaders to communicate in topical channels, and participants were able to mute channels they were not interested in as well as have private conversations. While this worked well for leaders and participants who have used Discord before, it did create a barrier to people who did not want to sign up for a separate service or were unable to check it regularly. Thus, the eIAYC organisers sent weekly emails to relevant mailing lists in addition to messaging in the Discord server. Overall, the eIAYC had 63 members as part of the eIAYC Discord server over the summer, roughly the size of a normal in-person camp.

In addition to announcing events via email and Discord, the eIAYC utilised Google calendar which was kept up to date throughout the summer. Participants were able to add it to their personal accounts and receive notifications.

### 4. Conclusions

Firstly, the eIAYC organisers significantly underestimated the amount of time it would take to plan the events. The constant expectation of having to be online every day for two and a half months added a lot of emotional pressure to the organisers. For the future, IWA has planned longer breaks for the eIAYC team.

Most importantly, IWA found that the event was too long: participation ebbed and flowed over the duration of the online camp. Unfortunately, towards the end of summer, scheduled events were often forgotten and attendance heavily depended on each participant's respective summer holidays. As a result, this year the eIAYC consists of a more limited set of higher impact events open to the public: (1) a seminar series consisting of seven one-hour sessions, happening every two weeks throughout the summer; (2) a 36-hour hackathon in July 2021; (3) and a virtual reunion weekend for former participants. This allows the eIAYC team to split into three small organising groups focused on each event, while reaching a larger audience, especially young people who have not participated in previous in-person IAYCs. Sign-ups to the eIAYC 2021 have reached more than 400 so far!

In addition, online events can be joined by anyone, regardless of barriers like money or visas, and the environmental impact is significantly less than the in-person IAYC. Based on these advantages it is worth maintaining online outreach activities post-pandemic.

Finally, at the end of the eIAYC 2020 the organisers conducted a feedback survey with 34 responses. While it was mentioned multiple times that it was difficult to interact with participants in other time zones and that it cannot replace the in-person experience, the eIAYC generally had good feedback. Asking the survey respondents to rank the overall eIAYC 2020 experience from 1 to 10 (with 10 being the highest), an average rating of 8.2 was achieved.